\documentclass[%
superscriptaddress,
reprint,
showpacs,preprintnumbers,
showkeys,
 amsmath,amssymb,
 aps,
prb,
]{revtex4-1}

\usepackage{graphicx}
\usepackage{epstopdf}
\usepackage{dcolumn}
\usepackage{bm}

\begin{document}


\title{Tunable Magnetic Domains and Depth Resolved Microstructure in Gd-Fe Thin Films}

\author{A. Talapatra}
\email{ph13p1001@iith.ac.in} \affiliation{Nanomagnetism and
Microscopy Laboratory, Department of Physics\\ Indian Institute of
Technology Hyderabad, Kandi, Sangareddy, 502285, Telangana,
India.}

\author{J. Arout Chelvane}
\affiliation{Defence Metallurgical Research Laboratory,
Hyderabad-500058, India.}

\author{B. Satpati}
\affiliation{Saha Institute of Nuclear Physics, HBNI, 1/AF
Bidhannagar, Kolkata 700064, India.}

\author{S. Kumar}
\affiliation{National Centre for Compositional Characterisation of
Materials, Hyderabad-500062, India.}

\author{J. Mohanty}
\email{jmohanty@iith.ac.in} \affiliation{Nanomagnetism and
Microscopy Laboratory, Department of Physics\\ Indian Institute of
Technology Hyderabad, Kandi, Sangareddy, 502285, Telangana,
India.}

\date{\today}

\begin{abstract}
This paper reports the tunability of magnetic domains and its
correlation with depth-resolved microstructure in 100~nm thick
Gd-Fe films after rapid thermal processing. The as-deposited films
are amorphous in nature and display elongated periodic stripe
domains, signifying the development of perpendicular magnetic
anisotropy. Rapid thermal processing at $550^oC$ for different
time intervals \textit{viz.}, 5, 10, 15 and 20~minutes induces the
crystallization of Fe over the amorphous Gd-Fe matrix and the
presence of wider reaction zone has been evidenced at the
film-substrate interface. While the magnetization measurements
reveal a spin re-orientation transition along the plane of the
film surface, the stripe domains are no longer observed for the
rapid thermal processed films. 3D micromagnetic simulations have
been performed to complement the experimental findings. The
consideration of in-plane tilt of anisotropy axis along with
anisotropy reduction successfully demonstrate the experimentally
observed changes in magnetic properties and domain structure.

\end{abstract}

\maketitle


\section{Introduction}
\label{sec1} Magnetic thin films have ushered in a new era in
magnetic and magneto-optical recording technology. With the rapid
research progress on magnetic materials with perpendicular
magnetic anisotropy (PMA), high density data storage ($1~Tb/in^2$)
is being attempted. Higher magnetic anisotropy makes the system
magnetically stable by minimizing the loss of information due to
thermal fluctuation of spins, which is technically known as
super-paramagnetism. However higher PMA is associated with higher
coercivity. This offers a new challenge for writing the media
within the limited field of the existing write heads. To
circumvent the problem, heat assisted magnetic recording has been
investigated, where the magnetic anisotropy is tailored to be
locally weak by laser heating to switch the media with a lower
external magnetic field. However, the implementation of this
method in a portable device is not very straightforward. So, the
search for different methods of anisotropy tuning by external
perturbations has gained an active interest in contemporary
research.

Gd-Fe is a popular material system due to its interesting property
of tunable PMA in amorphous as-deposited condition. It has been
the centre of attraction of the research community around the
globe since last few decades due to its applicability in magneto
optical recording. The reason behind the evolution of PMA and
reverse spin re-orientation transition (SRT) with respect to film
thickness and temperature is still debated. From the existing
literature, the contribution of the magnetic dipolar interaction
behind the determination of PMA can be
evidenced~\cite{MTC_JAP_79}. The role of structural parameters and
their correlations to the magnetic properties have been discussed
there. The surface SRT in Gd films on Fe in presence of magnetic
fields was elaborately explained by Camley exploiting mean field
theory and atomistic model~\cite{CRE_PRB_1987}. The phase
transition to a twisted spin state for anti-ferromagnetically
coupled Gd and Fe sublattices at higher field has also been
demonstrated by the same
researcher~\cite{CRE_PRB_1988,CRE_PRB_1989}. The generic sources
of PMA in amorphous rare earth-transition metal (RE-TM) alloy
systems have been pointed out in the light of fundamental
electro-magnetic theory~\cite{FMM_PRL_91}. The experimental
studies confirm that the sources are material specific and vary
with deposition process parameters~\cite{GRC_JVST_78}. From the
point of view of application in magneto-optical recording with
different RE-TM alloys, an excellent review has been provided by
Mieklejohn~\cite{MWH_IEEE_1986}, followed by Hansen \textit{et
al.} on magneto-optical properties of Gd, Tb, Fe, Co based binary
RE-TM alloys~\cite{HCM_JAP_1989}.

The interest shifted towards the domain imaging in Gd-Fe alloy or
in Gd/Fe multilayer system to a great extent in the beginning of
21st century. The high resolution domain imaging techniques along
with the study of material specific magnetic properties have been
developed with high intensity synchrotron X-ray. Eim\"{u}ller
\textit{et al.} utilized magnetic transmission X-ray microscopy
technique with X-ray magnetic circular dichroism for the
quantitative domain analysis in perpendicularly magnetized Gd/Fe
multilayer~\cite{EKF_JAP_2000}. A considerable piece of research
has been performed by Miguel and colleagues for the development of
X-ray resonant magnetic scattering to study the domain structure
and their dynamics in Gd-Fe thin
films~\cite{MPT_PRB_2006,Miguel_Thesis}. Utilizing the soft
ferrimagnetic nature of Gd-Fe, detailed study of exchange bias at
the soft-hard ferrimagnetic interfaces has been carried out by
Mangin {et al.}~\cite{MMS_PRB_2003,MHH_PRB_2006}. In the most
recent scenario, the RE-TM alloy thin films have gained interest
in the field of ultrafast magnetization dynamics due to the
different temporal response of the respective
sub-lattices~\cite{RSE_SPIN_15}. Gd/Fe multilayer system has also
been used for advanced dichroic coherent imaging
techniques~\cite{TMD_PNAS_2011}.

On the basis of the aforesaid existing research, we found that a
systematic study on magnetic domains and their modification in
Gd-Fe thin films has not been attempted yet. Konings \textit{et
al.} showed the anisotropy engineering in Gd-Fe thin films by
locking the domains into the pinning cites, created by focused ion
beam patterning~\cite{KML_JAP_2005}. But, the correlation with the
structural properties seems to be missing. However, tailoring of
magnetic anisotropy in TbFeCo thin films by rapid thermal
annealing has been reported by Umadevi {et
al.}~\cite{UBC_JAC_2016}. Lack of information on the magnetic
microstructure and its correlation with the magnetic properties
paved the motivation of our study. Here, we report the impact of
rapid thermal processing (RTP) on structural, micro-structural and
magnetic properties of electron beam (e-beam) evaporated Gd-Fe
thin films. Emphasis has been given towards different microscopy
techniques and complementary methods to understand the magnetic
microstructure in correlation with depth resolved microstructural
changes. The experimental results are complemented with
3D-micromagnetic simulation to observe the consequences of
anisotropy modification in magnetic thin films.

\section{Experimental Detail}
\label{sec2} 100 nm Gd-Fe films were deposited by e-beam
evaporation (Make: VST Israel) at room temperature on Si~$<100>$
substrates under background pressure of $2\times10^{-6}$ torr. The
deposition rate was optimized to be $0.2~nm/s$ employing an alloy
target with $Gd_{50}Fe_{50}$ composition. The substrate has been
rotated with a constant speed of 10 rpm to ensure uniformity in
thickness. A 3~nm thin layer of Cr was deposited as capping layer
to protect the films from oxidation. The thickness values
mentioned here are recorded using the quartz crystal monitor
associated to the deposition unit. RTP experiments have been
performed at $550^oC$ under high vacuum ($1\times10^{-7}$ torr)
for different time intervals \textit{viz.}, 5, 10, 15, 20 min with
a RTP furnace (Make: Milman Thin Film Pvt. Ltd., India), where the
ramp rate of heating was optimized at $250^oC/min$. The structural
information was obtained by grazing incidence X-ray diffraction
(GIXRD, Make: Bruker) using Cu $K_{\alpha}$ radiation at an angle
of incidence $1^o$. The microstructure of the films along the
cross-section and the film-substrate interface have been
characterized by cross-sectional transmission electron microscope
(XTEM), using FEI, Tecnai G2 F30, S-Twin microscope operating at
300~kV. Field emission scanning electron microscope (FESEM)
equipped with energy dispersive X-ray spectroscopy (EDS, Make:
Carl Zeiss) was used to observe the surface morphology along with
the estimation of overall composition. The depth resolved study of
composition was performed by Rutherford backscatterd spectrometry
(RBS) technique using carbon ($^{12}C$) beam of energy 6991~keV,
derived from a 3~MV tandetron accelerator (HVEE, Europa). The RBS
measurements were carried out in a vacuum of $5\times10^{-6}$ torr
and the backscatterd particles were detected at an angle of
$170^o$ using passivated implanted planar silicon (PIPS) detector.
The room temperature magnetization measurements have been carried
out with vibrating sample magnetometer (VSM) (Make: ADE
Technologies, USA, model EV7 VSM). Atomic force microscopy (AFM)
and magnetic force microscopy (MFM, Model: Bruker Dimension Icon)
were performed simultaneously by using CoCr coated antimony doped
Si tip at room temperature without any external magnetic field.


\section{Results and Discussions}
\label{sec3}

\begin{figure*}
\centering
\includegraphics[width=14cm, height=8.4cm]{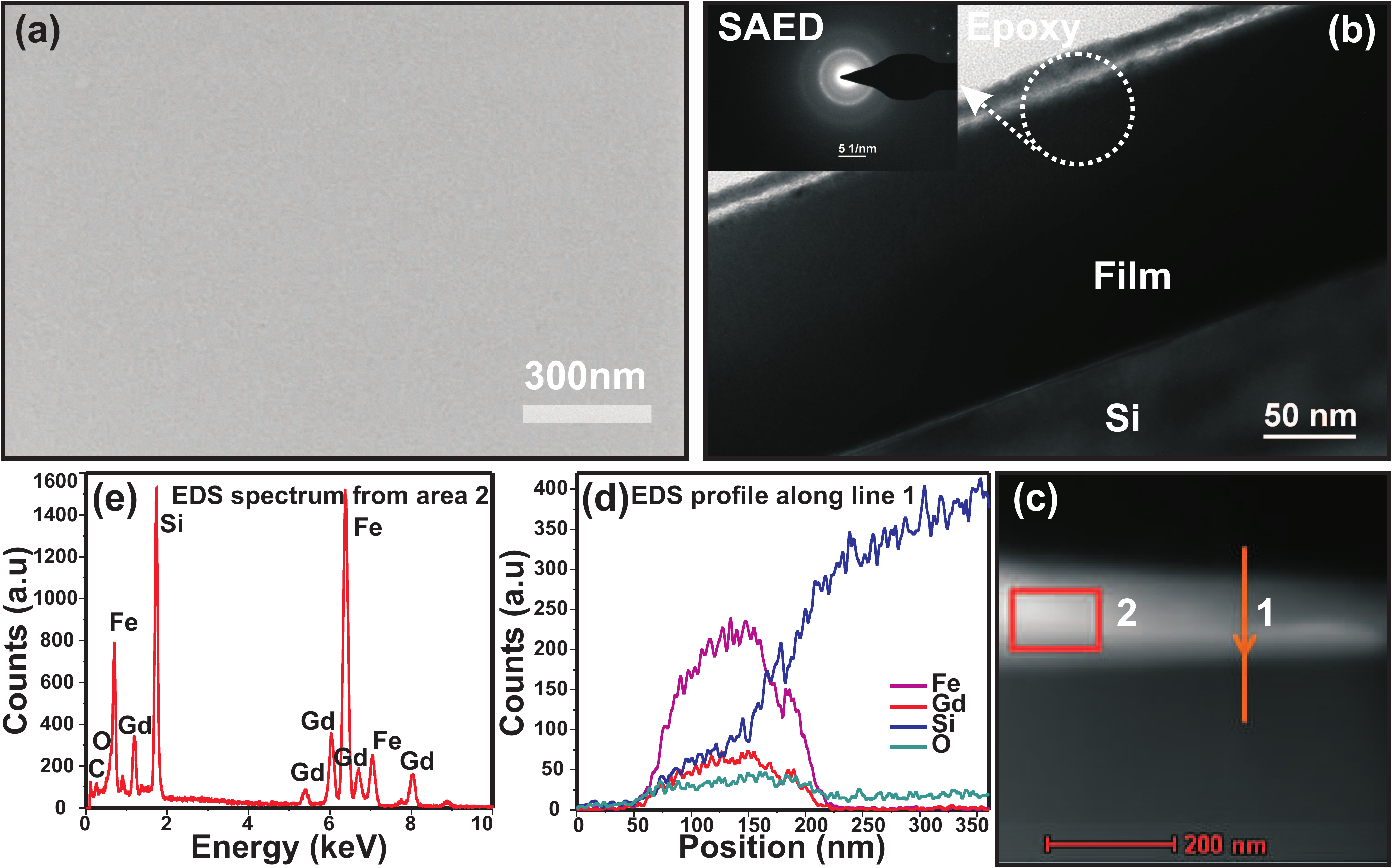}
\caption{(a) FESEM image, (b) XTEM image with SAED pattern as
inset for the as-prepared Gd-Fe thin film, (c) STEM-HAADF image,
(d) EDS line profile along line 1 in (c) and (e) EDS spectrum from
area 2 in (c).} \label{Fig.1}
\end{figure*}

The e-beam evaporated 100~nm thick Gd-Fe film shows a smooth
surface with no identifiable topographic features as in
Fig.~\ref{Fig.1}(a). The overall composition (atomic \%) of the
film has been found to be $Gd_{16.6}Fe_{83.4}$ using EDS,
associated with FESEM. The composition has been checked at
different places of the film and seen to be similar which
indicates the homogeneity in composition for the films. The
amorphous nature of the as-prepared film has been confirmed by the
selected area electron diffraction (SAED) pattern, derived from
the XTEM image, as shown in Fig.~\ref{Fig.1}(b). The
film-substrate interface is sharp, devoid of any diffusion or
intermixing. The EDS depth profile has been shown in
Fig.~\ref{Fig.1}(d), obtained across line 1 in the cross-sectional
high-angle annular dark-field scanning transmission electron
microscopy (HAADF-STEM) image (Fig.~\ref{Fig.1}(c)) and the
elemental composition is estimated to be $Gd_{13.7}Fe_{86.3}$
along the cross-section. The EDS spectrum from area 2 (in
Fig.~\ref{Fig.1}(c)) has also been depicted in
Fig.~\ref{Fig.1}(e). No other source of contamination has been
evidenced from the EDS spectrum, except the presence of oxygen.
For the as-prepared sample, the presence of oxygen has been
observed with increasing time, but the magnetic property does not
change considerably.

\begin{figure*}
\centering
\includegraphics[width=10.08cm, height=8cm]{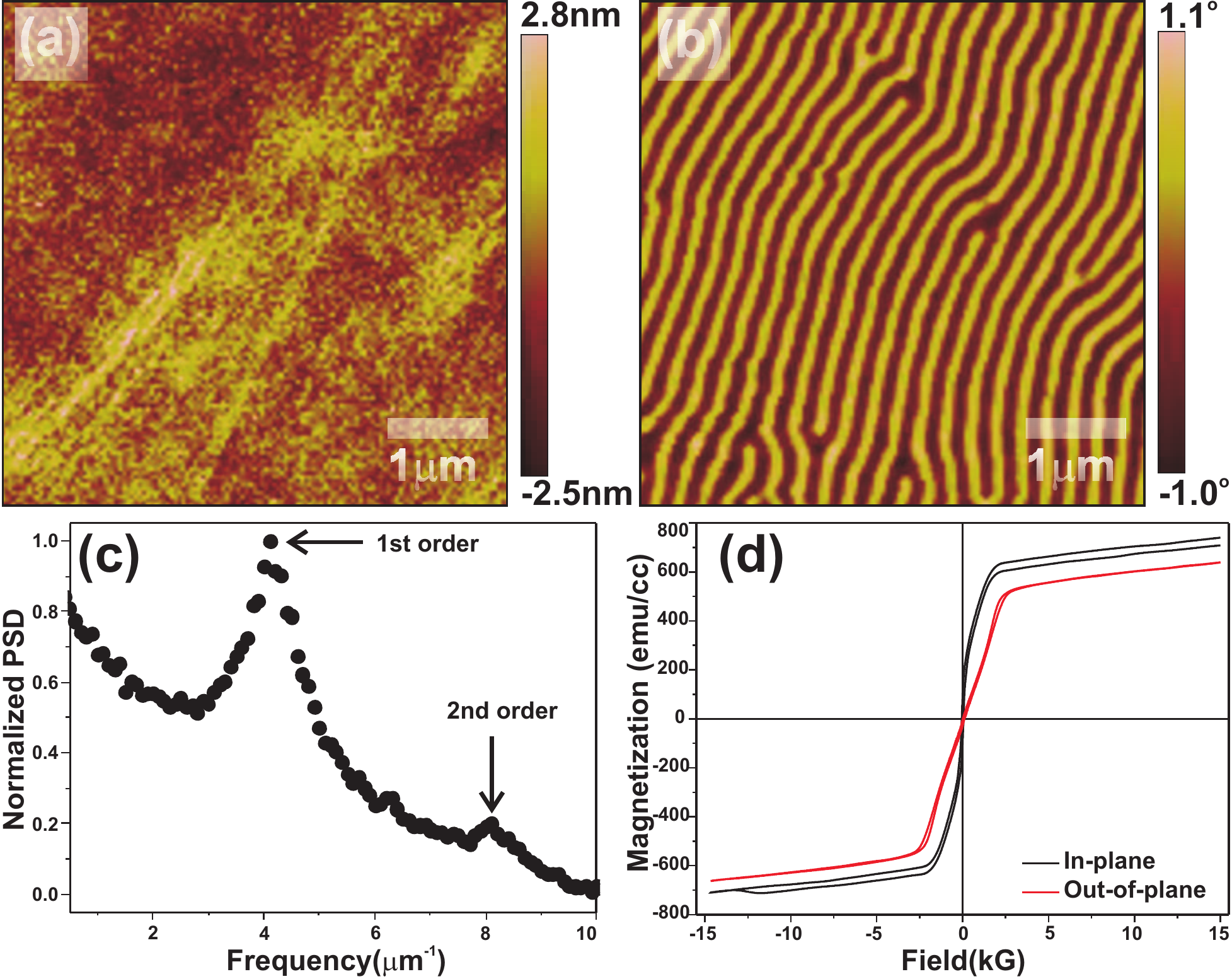}
\caption{(a) AFM, (b) MFM, (c) PSD profile of the MFM image and
(d) hysteresis loops along IP and OOP direction for 100~nm thick
as-prepared Gd-Fe thin film.} \label{Fig.2}
\end{figure*}

Fig.~\ref{Fig.2}(a) and (b) exhibit the AFM and MFM images of the
as-prepared film respectively, which indicate that the magnetic
phase is free from topographic influences. MFM reveals elongated,
periodic stripe domain pattern with high magnetic phase contrast.
The domains are essentially the self organization of magnetization
into alternating up-down magnetized areas with bright-dark
contrasts. They are formed due to the competition between the
exchange and anisotropy energies which favor a single domain state
with magnetization perpendicular to the film plane, with long
ranged demagnetizing field that favors the creation of in-plane
(IP) domain walls by breaking the system into multi-domain state.
The domain periodicity has been calculated from the 2D isotropic
power spectral density (PSD) profile as in Fig.~\ref{Fig.2}(c),
estimated from the Fourier transform of the MFM image. This allows
us to obtain the domain periodicity (twice the domain size) and
their relative weight~\cite{BMR_PRB_2007}. The peak of the PSD
curve points out the wavelength of the most predominant feature in
the image~\cite{BLW_APL_2011}. The PSD is fitted with a Lorentzian
to point out the mean value and standard deviation which account
for the domain periodicity and the associated error bar
respectively. The domain size turned out to be around 122 ($\pm
2$)~nm by inverting the frequency of the 1st order peak. The
presence of higher order peak signifies the long range correlation
of the domains. Similar kind of domains in as-prepared Gd-Fe films
has also been reported~\cite{MPT_PRB_2006}. But, the magnetic
properties are sensitive to the processing
conditions~\cite{BCR_TSF_2015,UCB_JMMM_2016} such as deposition
temperature, pressure, time and internal stress as well.
We have reported the evolution of meandering stripe domains in
Gd-Fe thin films and the enhancement of perpendicular contrast as
a function of film thickness by means of MFM and micromagetic
simulation~\cite{TAM_AIP_2017}. This kind of sub-200~nm sized
domains are usually the signature of the development of weak PMA
in thin films and results in a slanted hysteresis loop with almost
zero remanence when measured along the out-of-plane (OOP)
direction, as shown in Fig.~\ref{Fig.2}(d). Similar feature of
domains and hysteresis loop have also been reported for Co/Pt
multilayer~\cite{TAM_JMMM_2016,TAM_AIP_2016} and also in Gd-Fe
thin films~\cite{KML_JAP_2005}, where the presence of PMA has been
claimed for all the cases. While the coercivity for the IP (51
Gauss) and OOP (43 Gauss) loops are comparable, the remanence is
more for the IP loop as observed from Fig.~\ref{Fig.2}(d).

\begin{figure*}
\centering
\includegraphics[width=8.9cm, height=8cm]{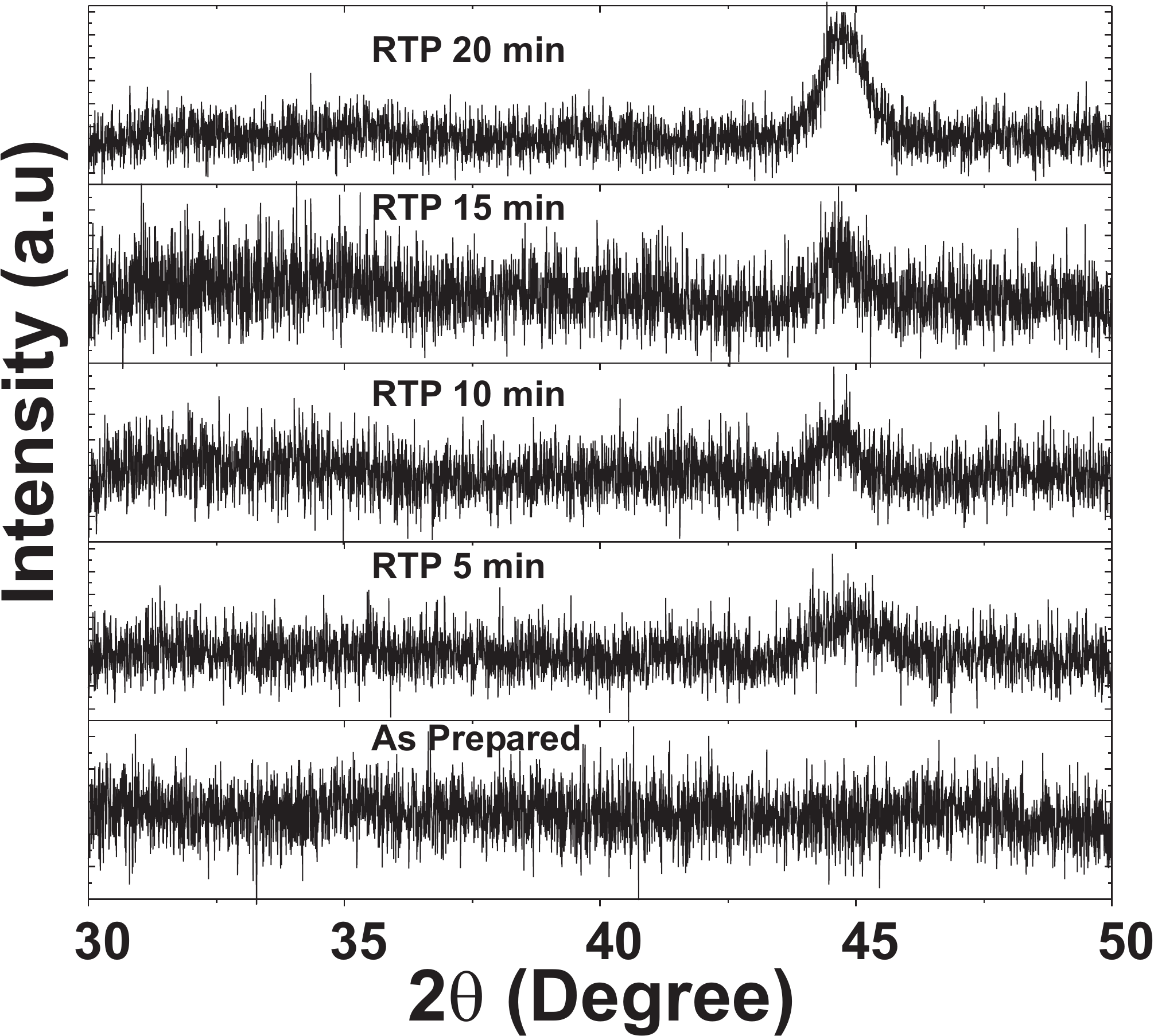}
\caption{ XRD spectra for the as-prepared Gd-Fe film and films
after RTP at $550^oC$ with different time intervals (mentioned in
respective places).} \label{Fig.3}
\end{figure*}

\begin{figure*}
\centering
\includegraphics[width=14cm, height=12.6cm]{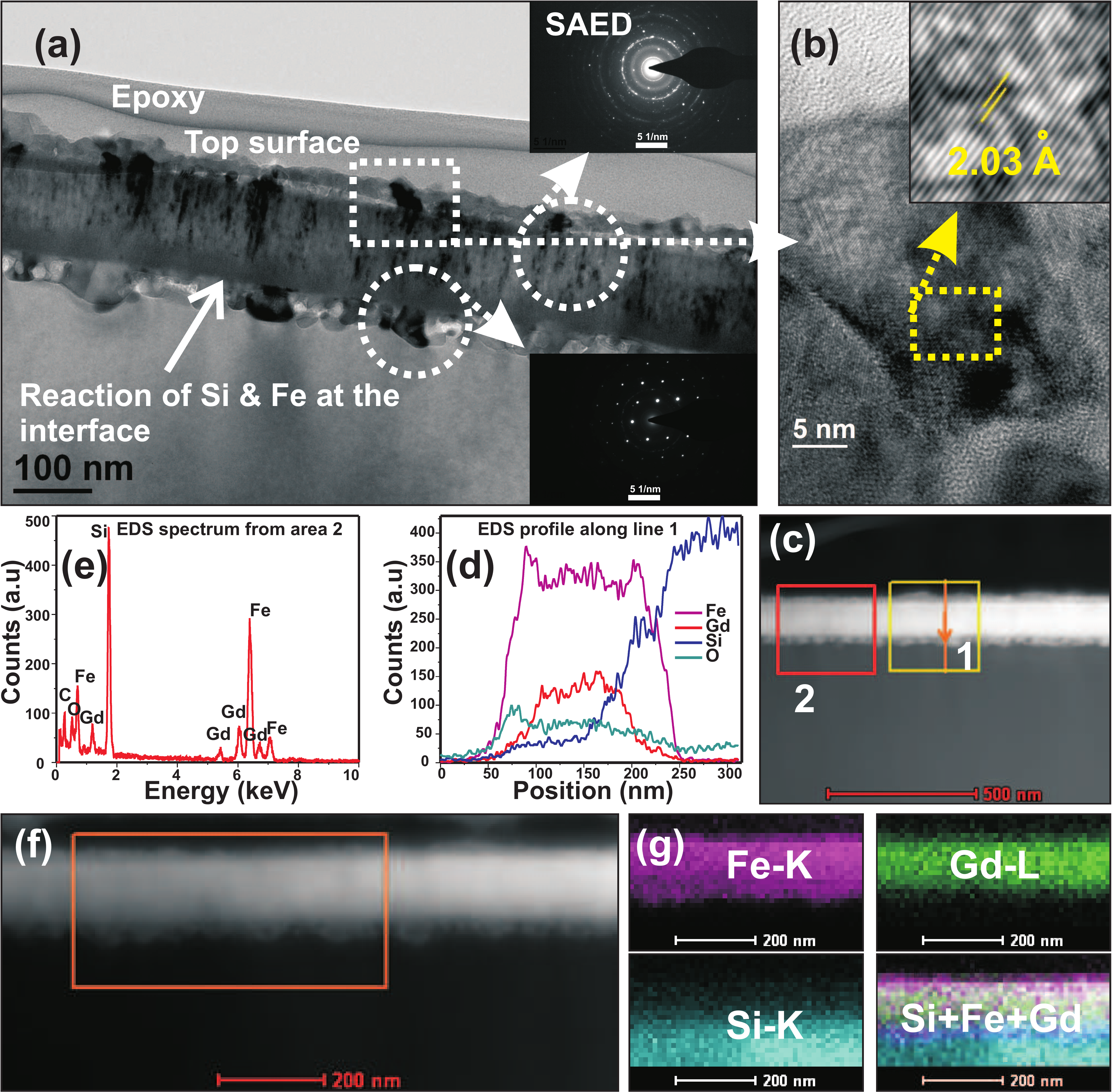}
\caption{ (a) XTEM image of Gd-Fe thin film after RTP for 20~min,
inset shows the SAED patterns taken from the top surface and at
the film-substrate interface, (b) HRTEM image of the selected area
from (a), inset represents IFFT image showing lattice fringes for
the specified area, (c) HAADF-STEM image, (d) EDS profile along
line 1 in (c), (e) EDS spectrum from area 2 in (c), (f) HAADF-STEM
image with (g) elemental mapping in the selected region.}
\label{Fig.4}
\end{figure*}

As our interest involves tailoring magnetic domains in Gd-Fe thin
films, so we have performed RTP at $550^oC$ for different time
intervals \textit{viz.} 5, 10, 15 and 20 min which results in a
global change in material properties, unlike local modifications
by laser~\cite{LMV_SCI_14,TAM_JMMM_2016} or focused ion beam
irradiation~\cite{MMK_PRB_2012,KML_JAP_2005} or recent trend of
voltage controlled magnetism~\cite{SCL_PMC_2017}. While the
as-prepared film shows amorphous nature, RTP induces crystallinity
in the films, degree of which increases upon increasing processing
time, as observed from the GIXRD data, shown in Fig.~\ref{Fig.3}.
The appearance of the peak at $2\theta \sim 44.67^o$ corresponds
to the highest intensity peak of Fe from (110) plane (JCPDS card
number: 00-006-0696). This designates the nucleation of
crystalline Fe over the amorphous Gd-Fe matrix.
Further, this has been verified by XTEM images, performed on the
film after RTP for 20~min (Fig.~\ref{Fig.4}(a)) where the
crystallinity is maximum. A clear indication about the presence of
Fe near the top and bottom surface has been evidenced from the EDS
line profile (Fig.~\ref{Fig.4}(d)) of the cross-sectional
HAADF-STEM image in Fig.~\ref{Fig.4}(c). The composition is
estimated to be $Gd_{15.4}Fe_{84.6}$ along the depth. The top
surface shows formation of nm-sized Fe crystallite, which is
evident from the SAED pattern (inset of Fig.~\ref{Fig.4}(a)). The
d-spacings calculated from the diameter of the rings match with
the reflections of Fe phase. The presence of spots confirms that
the Fe nanocrystals are distributed over the amorphous phase of
Gd-Fe. Furthermore, one of the crystalline Fe particles at the top
surface has been shown in the high resolution TEM (HRTEM) image of
Fig.~\ref{Fig.4}(b). The lattice spacing is estimated to be
2.03~\AA{} from the inverse Fourier filtered transform (IFFT)
image derived from the area under the yellow dotted box in the
HRTEM image and shown as the inset of Fig.~\ref{Fig.4}(b). The
value of the lattice spacing matches with (110) inter planar
spacing of Fe as per the mentioned JCPDS file. Similar observation
of the crystallization of FeCo alloy over amorphous TbFeCo matrix
has been reported~\cite{UBC_JAC_2016}. On top of that, elemental
mapping (shown in Fig.~\ref{Fig.4}(g))has been performed using
HAADF-STEM-EDS technique along the area shown in
Fig.~\ref{Fig.4}(f). This verifies the presence of Fe and Gd over
the considered area.

\begin{figure*}
\centering
\includegraphics[width=12cm, height=7.74cm]{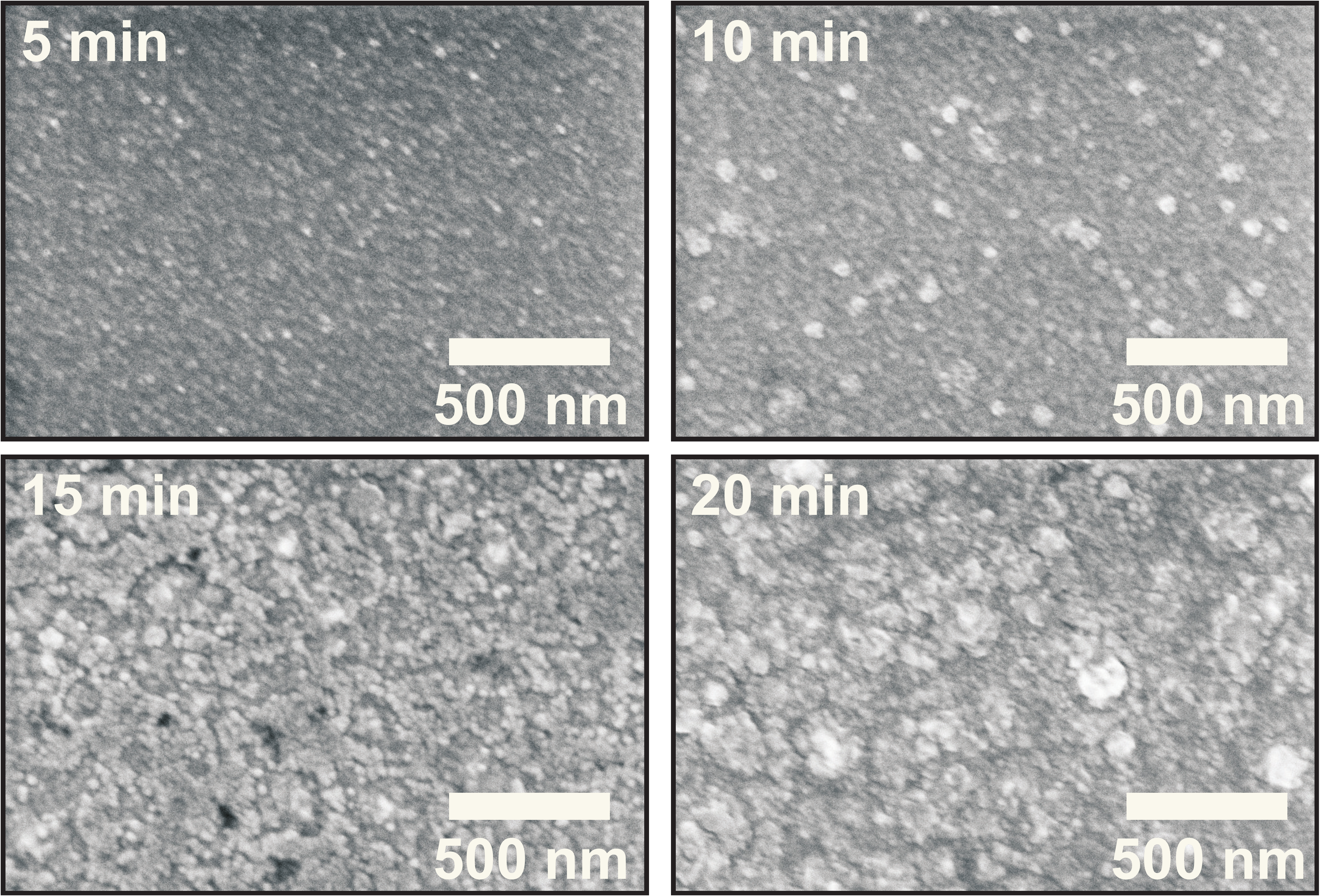}
\caption{FESEM surface micrographs for Gd-Fe films with different
RTP time intervals, mentioned in the top-left corner of the
images.} \label{Fig.5}
\end{figure*}

While the as-prepared film shows a feature-less smooth surface
(Fig.~\ref{Fig.1}(a)), the presence of island-like granular
microstructure has been evidenced in the FESEM surface
micrographs, as shown in Fig.~\ref{Fig.5}. It can be qualitatively
observed that the grains are randomly distributed over the surface
and the grain size increases with the increase in RTP time. At
20~min RTP time, the size becomes maximum with highest
irregularity in the distribution. The measurement of the overall
composition by EDS showed a very little increment of the
rare-earth metal upon increasing the RTP time and this is listed
in Table~\ref{table_1}.

\begin{figure*}
\centering
\includegraphics[width=12cm, height=9.375cm]{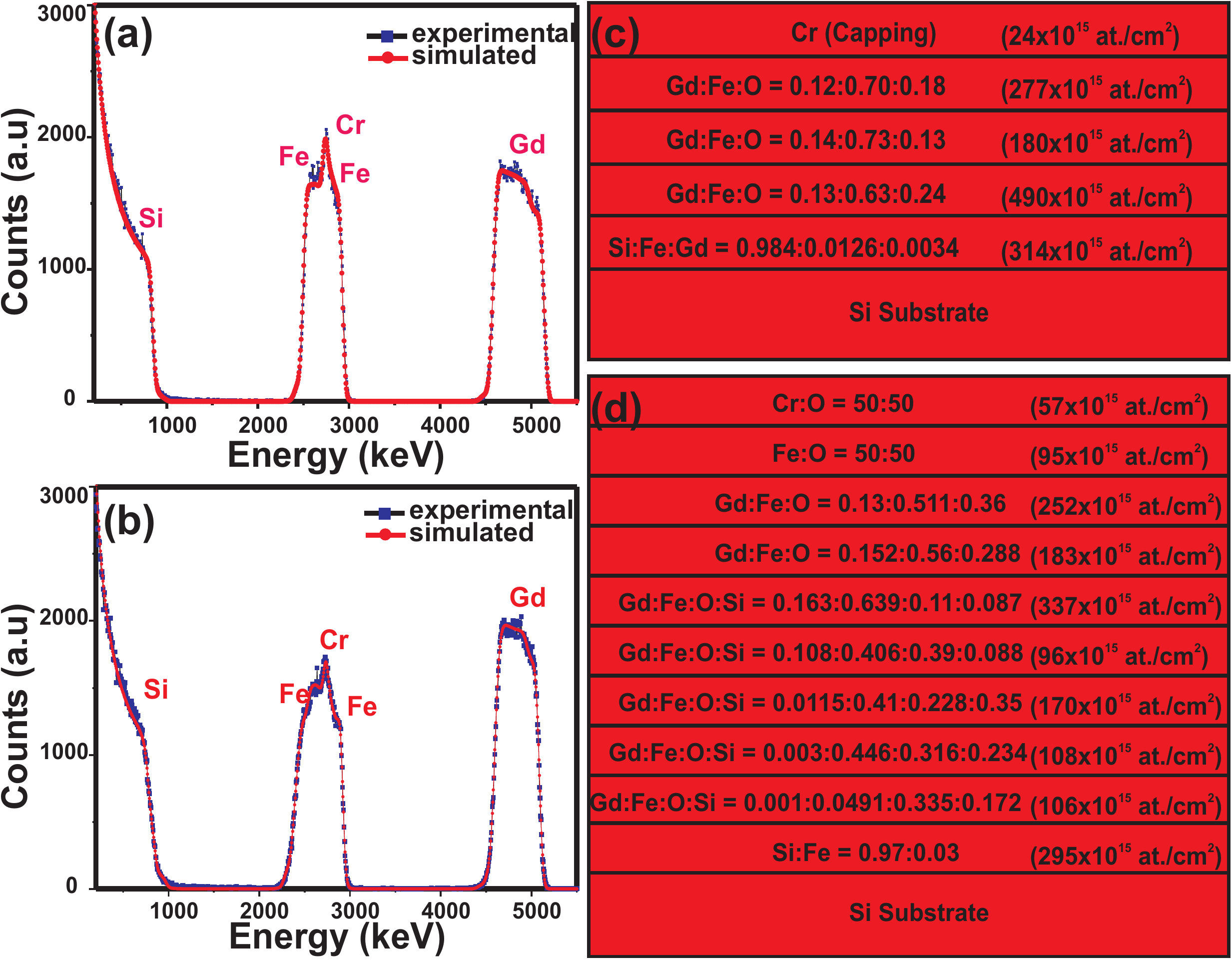}
\caption{ Experimental and simulated RBS spectra for (a)
as-prepared film and (b) the rapid thermal processed film for
10~min, (c) and (d) represent the simulation models for the case
of (a) and (b), respectively.} \label{Fig.6}
\end{figure*}

Further, the depth-resolved composition analysis has been carried
out with RBS technique. The experimental and the simulated RBS
spectra along with the simulation model have been shown in
Fig.~\ref{Fig.6}. The overall composition for the as-prepared film
has been estimated to be $Gd_{16.4}Fe_{83.6}$, from the area under
the experimental curve in Fig.~\ref{Fig.6}(a). This is in well
agreement with the EDS data. The best-fitted model of the
experimental data for the as-deposited film has been shown in
Fig.~\ref{Fig.6}(c), which shows the inhomogeneous distribution of
Gd and Fe across the cross-section of the film. The total
thickness of the film, estimated from RBS also matches within the
error bar of around 10\% with respect to the thickness, recorded
by the quartz crystal monitor. Now, the complexity arises after
performing RTP. The experimental curve in Fig.~\ref{Fig.6}(b)
shows the RBS spectra for the film with processing time of 10~
min. Calculation of the area under the curve unfurls that the
overall composition remains almost unchanged after RTP, which has
also been evident from the EDS data, recorded in
table~\ref{table_1}. The real picture of the depth-resolved
composition variation can be understood from the best-fitted
model, shown in Fig.~\ref{Fig.6}(d). The integrity of Cr capping
layer is commendable as it does not show any diffusion even after
RTP. The most important feature is the nucleation of Fe,
predominantly taking place near the top surface, elucidated from
the RBS study. This is also in concurrence with the XTEM results,
shown in Fig.~\ref{Fig.4}(a). The reason behind the nucleation of
Fe  below the Cr layer could be probably due to the fact that Cr
acts as nucleation center for the formation of Fe nano-crystal.
This can be attributed to the closeness of the crystal structure
and lattice parameter of Fe and Cr in comparison with Gd-Fe and
Fe. On the other hand, the diffusion of Si has been evident near
to the film-substrate interface which increases the probability
for the formation of stable iron-silicide alloy at the interface.
Moreover, the diffusion aided kinetics makes the reaction zone
wider with the increase in RTP time~\cite{UBC_JAC_2016}. Thus, the
nucleation of Fe near the top surface and the formation of wider
reaction zone at the film substrate interface results in the
ingression of Gd and hence the presence of amorphous Gd-Fe alloy
has been evidenced only in few intermediate layers. The results of
RBS experiments are found to be in good agreement with the XTEM
results (Fig.~\ref{Fig.4}), where we get an extra information
about the crystallization of Fe.

\begin{figure*}
\centering
\includegraphics[width=14cm, height=5.5cm]{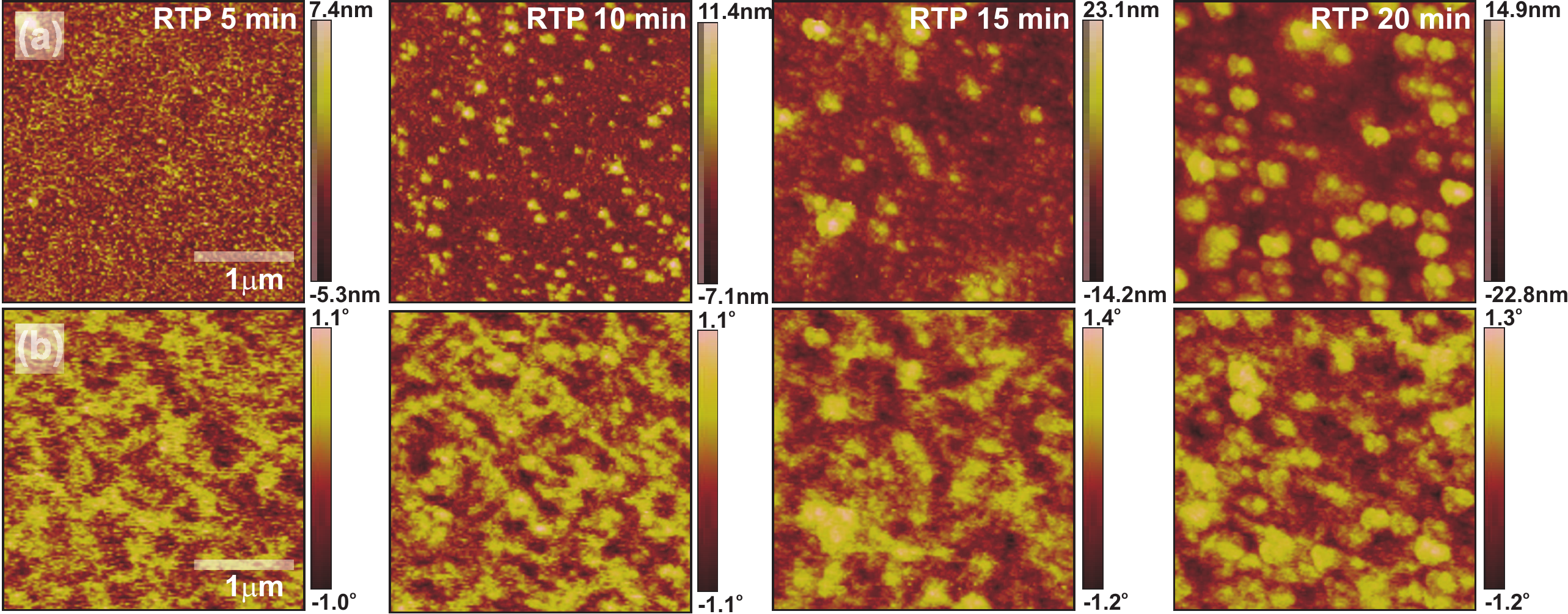}
\caption{ (a) AFM and (b) MFM images for the rapid thermal
processed films with different time intervals, mentioned in the
top-right corner of the respective AFM images. The area of
scanning for the AFM and MFM images is same (vertically). The
scale bar is same for all the images.} \label{Fig.7}
\end{figure*}

To probe the topography microscopically with higher resolution,
AFM measurement has been carried out for all the films and are
shown in the series of Fig.~\ref{Fig.7}(a). Similar findings of
increasing nano-crystalline grain growth has also been noticed
with increasing RTP time. To get quantitative information from the
AFM images, root mean square roughness ($R_q$), grain diameter and
the most probable threshold height ($H_{Th}$) of the topographic
features have been estimated and mentioned in Table~\ref{table_1}.
It has been observed that the $R_q$ increases monotonically with
RTP time and reaches almost 4.25 times when RTP is performed for
20~min compared to that for the duration of 5~min. Due to the
increasing RTP time, the crystalline grain diameter increases, but
the size distribution gets broader (as the error is very high) for
20~min sample. This can be attributed to the increase in
non-uniformity in the distribution of the crystallites over the
film surface which has also been qualitatively understood from the
FESEM images (Fig.~\ref{Fig.5}). From the point of view of
scanning probe microscopy, $H_{Th}$ is an important criterion
which designates the average height of the topographic feature
with respect to the maximum height in the depth histogram
analysis. The $H_{Th}$ turns out to be maximum for the film with
processing time of 15~min and again drops down for the film
processed for 20~min. Comparatively higher value of the full width
half maxima (FWHM) essentially proves again the non-uniform growth
of the crystallites for the films with maximum processing time.

\begin{table*}[ht]
 \caption{Variation of overall elemental composition, $R_q$, grain size and $H_{Th}$ of grains with RTP time interval.}
 \centering
 \begin{tabular}{c c c c c}
 \hline
 RTP Time      & Composition            & $R_q$      & Grain size ($\pm$ error)   & $H_{Th}$ ($\pm$ FWHM) \\
 (min)         & (atomic \%)            & (nm)       & (nm)                       & (nm)\\
 \hline
  5 min   & $Gd_{16.9}Fe_{83.1}$   & 2.0       & 140.4 ($\pm$21.6)            & 21.3 ($\pm$3.2)\\
  10 min  & $Gd_{17.2}Fe_{82.8}$   & 2.5       & 138.4 ($\pm$34.3)            & 43.3 ($\pm$3.4)\\
  15 min  & $Gd_{18.1}Fe_{81.9}$   & 4.8       & 173.5 ($\pm$74.4)            & 107.2 ($\pm$7.2)\\
  20 min  & $Gd_{18.1}Fe_{81.9}$   & 8.3       & 289.0 ($\pm$205.8)           & 49.9 ($\pm$15.1)\\[1ex]
 \hline
  \end{tabular}
\label{table_1}
\end{table*}

Magnetic microstructure of the films has been imaged with MFM
using a magnetized tip with a tip lift height of 50~nm from the
film surface and the images are shown in Fig.~\ref{Fig.7}(b). The
area of scanning for the AFM and MFM images are same which helps
to correlate the magnetic microstructure with the topographic
feature. While the as-prepared film displays elongated stripe
domains (Fig.~\ref{Fig.2}(b)), the film with processing time of
5~min does not show any identifiable domain pattern. Two different
magnetic contrasts appear in the form of irregular patches which
essentially means that the PMA of the sample gets weaker and the
formation of stripe domains is no longer energetically favorable.
This may take place due to the modification of magnetic anisotropy
after the RTP treatment. Hence, the magnetic phase contrast is
expected to be weaker for the films with higher RTP time. We are
unable to visualize this from the MFM images because the
topographic features dominate over the magnetic phase. Starting
from the sample with RTP time of 10~min, the $H_{Th}$ of the
crystallites (Table~\ref{table_1}) becomes comparable to the lift
height of the MFM tip. As a result, short ranged Van der Waals
force wins over the long ranged magnetic force and topography
dominated mixed phase is obtained. MFM has also been performed
with higher lift heights (not shown here), but the weaker magnetic
signal with strong topography results in noise in the images.

\begin{figure*}
\centering
\includegraphics[width=12cm, height=7.5cm]{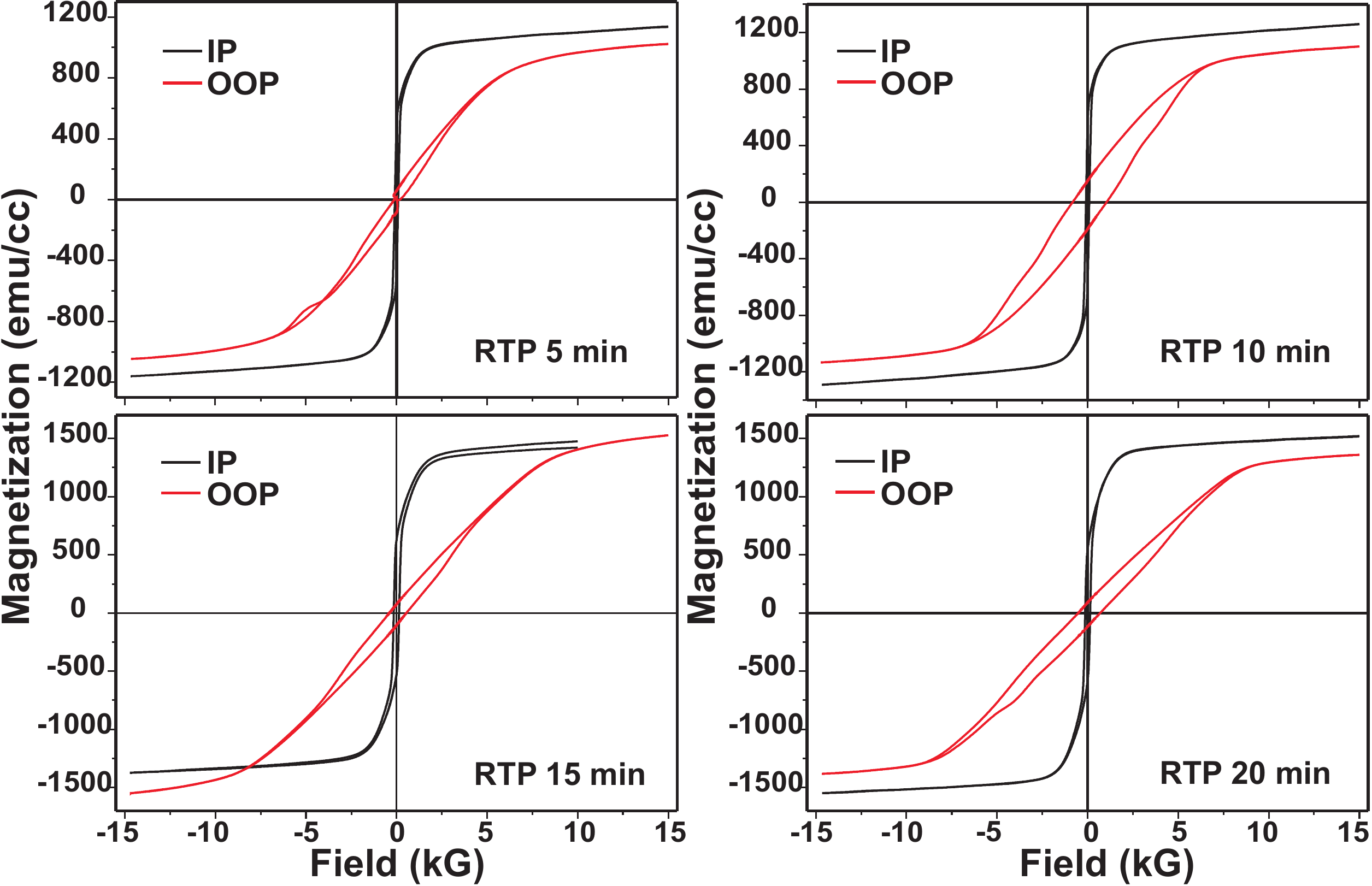}
\caption{ IP and OOP hysteresis loops for the films, processed
with different RTP time, mentioned in the bottom-right corner of
the respective images.} \label{Fig.8}
\end{figure*}

Fig.~\ref{Fig.8} represents the hysteresis loops for the films
after RTP, measured along the IP and OOP geometry. The extracted
magnetic properties are summarized in Table~\ref{table_2}. It can
be observed that the remanence and saturation magnetization
($M_s$) shows increasing trend with increasing RTP time for both
the IP and OOP mode. But the specific value of the IP remanence is
much greater than that in OOP mode. The increase in squareness
(ratio of remanence to saturation magnetization) for the IP loops
with RTP time suggests that RTP favors the magnetization to lie in
the plane of the film surface, unlike in the case of the
as-prepared film. Umadevi \textit{et al.}~\cite{UBC_JAC_2016}
calculated the tilt angle of hard Tb-Fe-Co with increasing
annealing time by using a formula mentioned in the
reference~\cite{CMF_JPD_2013} employing cosine inverse of the
ratio of the remanent magnetization for the film after RTP to that
of the as-prepared film. This calculation may show some meaningful
result for the case of materials with strong PMA as in the case of
Tb-Fe-Co, where the as-prepared film shows almost square
hysteresis loop. Gd-Fe with comparably weak PMA, the way of
calculation of tilt angle by similar method does not hold good. We
need to understand the anisotropy modification in terms of
increase in IP remanence and squareness compared to the OOP mode.
On top of that, the OOP saturation field is almost three times the
IP saturation field, which indicates that the IP direction becomes
the preferential orientation of magnetization after RTP.

\begin{table*}[ht]
 \caption{Variation of IP and OOP Remanence ($M_R$), coercivity ($H_c$), saturation magnetization ($M_s$) and saturation field ($H_s$) with RTP time interval.}
 \centering
 \begin{tabular}{c c c c c}
 \hline
 RTP Time      & $M_R$ IP (OOP)  & $H_c$ IP (OOP) & $M_s$ IP (OOP) & $H_s$ IP (OOP) \\
 (min)         & (emu/cc)        & (Gauss)        & (emu/cc)       & (Gauss)\\
 \hline
 5   & 462 (70)        & 97 (118)       & 1007 (862)     & 2138 (6585)\\
 10  & 559 (171)       & 92 (938)       & 1138 (1002)    & 3283 (7445)\\
 15  & 579 (89)        & 152 (474)      & 1260 (1404)    & 3292 (9988)\\
 20  & 562 (102)       & 146 (609)      & 1414 (1293)    & 3345 (10020)\\[1ex]
 \hline
  \end{tabular}
\label{table_2}
\end{table*}

It is well known that, the magnetic moment of Fe decreases when
alloyed with rare-earth materials because of the crystal field
effect. It has been reported that the magnetic moment of Fe in
$GdFe_2$ is estimated to be around $1.85\mu_B$ (by M\"{o}ssbauer
studies), whereas the magnetic moment of free Fe is
$2.2\mu_B$~\cite{AMM_HI_2008}. Thus the nucleation of free Fe
makes the $M_s$ of the processed films higher in comparison to the
as-prepared film. Now, $M_s$ and effective anisotropy $(K_{eff})$
are related as:

\begin{equation}
\centering
 K_{eff} = K_u - \frac{1}{2} \mu_0 M_s^2
\end{equation}
\label{equation_1}

where $K_u$ is the uniaxial anisotropy and $\mu_0$ is the magnetic
permeability in free space. Hence, it is clear from the equation
that increase in $M_s$ makes the effective perpendicular
anisotropy lower. On top of that, Fe shows only in-plane
anisotropy except for the case of controlled monolayer growth.
Thus the observed magnetization behavior can be attributed to the
Fe crystallization in the amorphous Gd-Fe matrix, which is
responsible for the SRT from OOP to IP direction after RTP.
However, the coercivity value is observed to be more in the OOP
mode and varies somewhat randomly which can be corroborated to the
increase in pinning sites along vertical direction due to the
increased $H_{Th}$ of the nano-crystallites.

In the light of the magnetic microstructure and magnetization
measurements, it is evident that the magnetic anisotropy is
getting modified due to RTP. But the experimental limitation
hindered us to image the magnetic domain structure independently,
without topographic influence. This has been complemented by 3D
micromagnetic simulation, which directly shows the variation of
magnetic domains as a function of anisotropy value as well as its
IP tilt along the $xz$ plane. The simulations of magnetic domains
are carried out by OOMMF software~\cite{OOMMF} which performs the
minimization of the total magnetic energy of the system, starting
from a random state with uniform vertical ($+Z$) magnetization.
The input parameters used for the simulations have been taken from
literature~\cite{MPT_PRB_2006} which include $M_s$ = $221~emu/cc$,
$K_u$ = $180~kerg/cc$ and the exchange stiffness constant ($A$) =
$2.6 \times 10^{-7}~erg/cm$. The volume of each cell is taken to
be $(4~nm)^3$, which is smaller than the exchange length
($\sqrt{2A/\mu_0.M_{s}^{2}} \sim 9~nm$) of the material. The
effect of finite temperature has not been considered in the
simulations.

\begin{figure*}
\centering
\includegraphics[width=12cm, height=5.4cm]{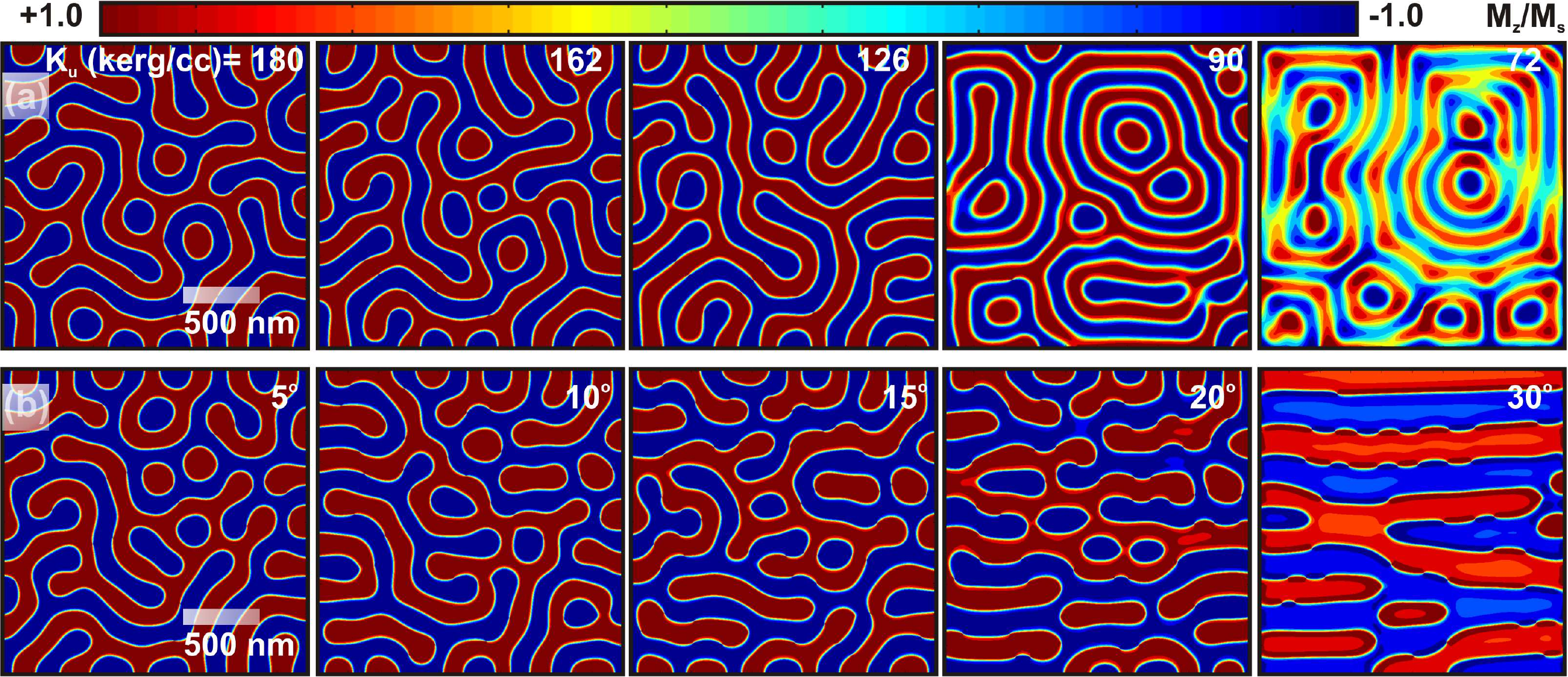}
\caption{ Simulated domain images with (a) reduced anisotropy
values and (b) with IP tilt along $xz$ plane. The anisotropy value
(in $kerg/cc$) and the value of the tilt angle (in degree) have
been mentioned at the top-right corner of the respective images in
the series of (a) and (b) respectively. The scale bar and color
bar are same for all the images.} \label{Fig.9}
\end{figure*}

\begin{figure*}
\centering
\includegraphics[width=12cm, height=4.7cm]{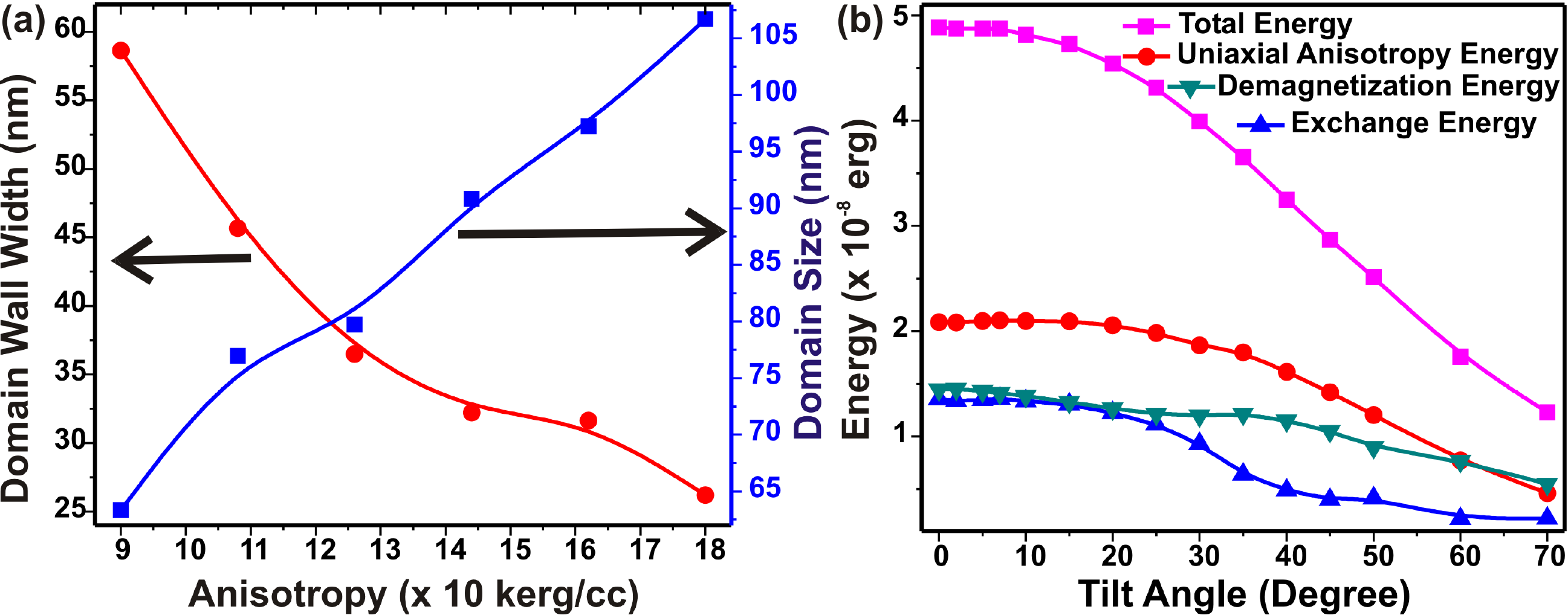}
\caption{ Variation of (a) domain size and domain wall width as a
function of anisotropy value, (b) different micromagnetic energy
terms along with total energy as a function of anisotropy tilt
angle.} \label{Fig.10}
\end{figure*}

The first image of Fig.~\ref{Fig.9}(a) shows the domain structure
($M_z$ component of the magnetization vector) for a 100~nm thick
Gd-Fe film, specified by the aforementioned input parameters. The
two different contrasts (red and blue) in the image represent
oppositely magnetized alternating up-down domains having
magnetization parallel or anti-parallel to the surface normal. The
total area of simulation ($2\mu m \times 2\mu m$) is much greater
than the characteristic feature size of the domains, so the effect
of truncation can be neglected. The domain in the first image of
Fig.~\ref{Fig.9}(a) shows a perfect demagnetized state with
connected stripes and that can be considered to be comparable to
the domain pattern for the as-prepared film as shown in
Fig.~\ref{Fig.2}(b). Line scans over the different areas of the
domains indicate the average domain size to be around 107 ($\pm$
5) nm, which is comparable to the domain size obtained from the
experiment. On top of that, the domain wall width
($\sqrt{M_{x}^{2} + M_{y}^{2} }$ component) turned out to be 26
($\pm$ 2) nm which matches very close to the theoretical value of
28~nm ($\pi.\sqrt{A/K_u}$). In order to study the effect of
anisotropy in OOP to IP transition of the magnetization, we have
systematically decreased the anisotropy value at an interval of
10\% of the original $K_u$ value, mentioned earlier. The variation
of domain size and domain wall width as a function of anisotropy
value has been depicted in Fig.~\ref{Fig.10}(a).

The simulated domain size decreases with decreasing the anisotropy
and simultaneously the domain wall width increases. Finally, at
$K_u = 90~kerg/cc$ (50\% less compared to the original $K_u$), IP
components come into the picture and as a result of that the
elongated stripe pattern tries to shrink to a closed circular-like
domains to minimize the magnetostatic energy. The comparable
values of domain size and domain wall width at the same anisotropy
value (Fig.~\ref{Fig.10}(a)) also signifies a strong competition
between the IP and OOP components of the magnetization. Similar
observations on domain pattern were also reported for Co/Pt
multilayer in response to ultrafast laser pulse and was also
verified with micromagnetic simulations~\cite{TAM_JMMM_2016}. The
series of images in Fig.~\ref{Fig.9}(b) shows the effect of
anisotropy tilt along the $xz$ plane. The first image in
Fig.~\ref{Fig.9}(b) series shows the domain configuration at $5^o$
tilt angle with respect to the fundamental one, shown in the first
image of Fig.~\ref{Fig.9}(a). With increasing tilt angle, the
domain structure is getting modified and the random stripe domain
pattern is converted to irregular discrete bubbles and patches
with a preferential orientation along the longitudinal direction.
This effect is very clear from the domain image with tilt angle of
$20^o$ and more. Moreover, the contrast variation in the
normalized images confirms the reduction in PMA and the resulting
anisotropy does not energetically favor the formation of elongated
periodic stripe domains which has also been evidenced from the
experimental results. Recently we have reported a systematic
transformation of magnetic domains from a meandering stripe state
to elongated stripes with preferred orientation which further gets
converted to magnetic patches (no-stripe state) with the increase
in RTP temperature~\cite{TAM_JMMM_2018}. While the irregular
domain pattern with different color contrast hinders to estimate
the domain size properly, the energetics of the system
(Fig.~\ref{Fig.10}(b)) clearly throw light towards the
micromagnetic phenomena. The monotonic decrease in total energy of
the system is followed by the decrease in other magnetic energies,
namely uniaxial anisotropy energy ($E_K$), exchange energy ($E_A$)
and demagnetization energy ($E_D$) with increasing tilt angle. But
the maximum variation of $E_K$ is calculated to be 75\% whereas
the other two energies show a negligible variation up to the tilt
angle of $30^o$. This may be considered as the threshold tilt
angle in this material system to offer a SRT after which IP
component of magnetization becomes dominant. As a result, $E_A$
and $E_D$ drops down to 84\% and 62\% respectively. Thus, our
simulation results correlate the effect of magnetization tilt in
magnetic microstructure as well as the energetics of the system.

\section{Conclusion}
\label{sec4} Detailed study on structure, microstructure and
magnetic properties have been performed on e-beam evaporated
100~nm Gd-Fe films, subjected to RTP for different time intervals.
The as-deposited films are amorphous and display elongated stripe
domains with higher magnetic phase contrast which symbolizes the
development of PMA in the material. RTP treatment at $550^oC$ for
different time intervals induces the nucleation of Fe nanocrystals
over the amorphous Gd-Fe matrix. The detailed structural study by
XTEM reveals the presence of nanocrystalline Fe on the top surface
and diffusion of Si at the film-substrate interface,
quantification of which has been provided by RBS. FESEM surface
micrographs and AFM show the irregular distribution of island-like
grains on the surface, characteristic size of which increases with
increasing RTP time. The magnetization studies explained that the
films underwent a SRT with magnetization lying in the plane of the
film surface upon different time interval of RTP, which could be
probably due to the crystallization of Fe that gives an IP tilt to
the magnetic moment. Special thrust has been given to understand
the magnetic microstructure with the increase in RTP time. The
decrease in magnetic contrast and simultaneous growth of
topographic feature limit us to get a topography free magnetic
contrast by MFM. To unravel the modification of domain structure,
3D micromagnetic simulations have been carried out which
successfully explain the change in the domain pattern, their size,
orientation and micromagnetic energetics of the system under the
modification of anisotropy value and its direction.

\section*{Acknowledgment}
\label{sec5}

The complete work has been carried out under the scope of the
project DMRL/O/CARS-17, issued by DRDO, India.

\bibliographystyle{elsarticle-num}


\begin{thebibliography}{26}

\bibitem{MTC_JAP_79}
T. Mizoguchi, G. S. Cargill, Magnetic anisotropy from dipolar
interactions in amorphous ferrimagnetic alloys, J. Appl. Phys. 50
(1979) 3570.

\bibitem{CRE_PRB_1987}
R. E. Camley, Surface spin reorientation in thin Gd films on Fe in
an applied magnetic field, Phys. Rev. B 35 (7) (1987) 3608.

\bibitem{CRE_PRB_1988}
R. E. Camley, D. R. Tilley, Phase transitions in magnetic
superlattices, Phys. Rev. B 37 (7) (1988) 3413.

\bibitem{CRE_PRB_1989}
R. E. Camley, Properties of magnetic superlattices with
antiferromagnetic interfacial coupling: magnetization,
susceptibility, and compensation points, Phys. Rev. B 39 (16)
(1989) 12316.

\bibitem{FMM_PRL_91}
H. Fu, M. Mansuripur, P. Meystre, Generic source of perpendicular
magnetic anisotropy in amorphous rare-earth-transition-metal
films, Phys. Rev. Lett. 66 (8) (1991) 1086.

\bibitem{GRC_JVST_78}
R. J. Gambino, J. J. Cuomo, Selective resputtering induced
anisotropy in amorphous films, J. Vac. Sci. Tech. 15 (1978) 296.

\bibitem{MWH_IEEE_1986}
W. H. Meiklejohn, Magnetooptics: a thermomagnetic recording
technology, Proc. IEEE 74 (11), (1986) 1570.

\bibitem{HCM_JAP_1989}
 P. Hansen, C. Clausen, G. Much, M. Rosenkranz, K. Witter, Magnetic and magneto-optical properties of rare-earth transition-metal alloys containing Gd, Tb, Fe, Co, J. Appl. Phys. 66 (1989) 756.

\bibitem{EKF_JAP_2000}
T. Eim\"{u}ller, R. Kalchgruber, P. Fischer, G. Sch\"{u}tz, P.
Guttmann, G. Schmahl, M. K\"{o}hler, K. Pr\"{u}gl, M. Scholz, F.
Bammes, G. Bayreuther, Quantitative imaging of magnetization
reversal in FeGd multilayers by magnetic transmission x-ray
microscopy, J. Appl. Phys. 87 (9) (2000) 6478.

\bibitem{MPT_PRB_2006}
 J. Miguel, J. F. Peters, O. M. Toulemonde, S. S. Dhesi, N. B. Brookes, J. B. Goedkoop, X-ray resonant magnetic scattering study of magnetic stripe domains in $\alpha$-GdFe thin films, Phys. Rev. B 74 (2006) 094437.

\bibitem{Miguel_Thesis}
 Miguel, J. Ph.D thesis, \textquotedblleft Static and dynamic X-ray resonant magnetic scattering studies
on magnetic domains\textquotedblright, University of Amsterdam
(2005),
\url{https://iop.fnwi.uva.nl/cmp//scientific_output/phd_theses.html}.

\bibitem{MMS_PRB_2003}
 S. Mangin, F. Montaigne, A. Schuhl, Interface domain wall and exchange bias phenomena in ferrimagnetic/ferrimagnetic bilayers, Phys. Rev. B 68 (2003) 140404(R).

\bibitem{MHH_PRB_2006}
 S. Mangin, T. Hauet, Y. Henry, F. Montaigne, E. E. Fullerton, Influence of lateral domains and interface domain walls on exchange-bias phenomena in GdFe / TbFe bilayers, Phys. Rev. B 74 (2006) 024414.

\bibitem{RSE_SPIN_15}
 I. Radu, C. Stamm, A. Eschenlohr, F. Radu, R. Abrudan, K. Vahaplar, T. Kachel,
 N. Pontius, R. Mitzner, K. Holldack, A. F\"{o}hlisch, T. A. Ostler, J. H. Mentink,
 R. F. L. Evans, R. W. Chantrell, A. Sukamoto, A. Itoh, A. Kirilyuk, A. V. Kimel, T. Raising, Ultrafast and distinct spin dynamics in magnetic alloys, Spin 5(3) (2015) 1550004.

 \bibitem{TMD_PNAS_2011}
A. Tripathi, J. Mohanty, S. H. Dietze, O. G. Shpyrko, E. Shipton,
E. E. Fullerton, S. S. Kim, I. Mcnulty, Dichroic coherent
diffractive imaging, Proc. Nat. Acad. Sci. 108 (33) (2011) 13393.

\bibitem{KML_JAP_2005}
 S. Konings, J. Miguel, J. Luigjes, H. Schlatter, H. Luigjes, J. Goedkoop, V. Gadgil, Lock in of magnetic stripe domains to pinning lattices produced by focused ion-beam patterning, J. Appl. Phys. 98 (2005) 054306.

\bibitem{UBC_JAC_2016}
 K. Umadevi, S. Bysakh, J. Arout Chelvane, S. V. Kamat, V. Jayalakshmi, Tailoring magnetic anisotropy in Tb-Fe-Co thin films by rapid thermal annealing, J. Alloy. Comp. 663 (2016) 430.

\bibitem{BMR_PRB_2007}
 V. Baltz, A. Marty, B. Rodmacq, B. Dieny, Magnetic domain replication in interacting bilayers with out-of-plane anisotropy: application to Co/Pt multilayers, Phys. Rev. B 75 (2007) 014406.

\bibitem{BLW_APL_2011}
 S. R. Bakaul, W. Lin, T. Wu, Evolution of magnetic bubble domains in manganite films, Appl. Phys. Lett. 99 (2011) 042503.

\bibitem{BCR_TSF_2015}
 H. Basumatary, J. Arout Chelvane, D. V. Sridhar Rao, S. V. Kamat, R. Ranjan, Effect of sputtering parameters on the structure, microstructure and magnetic properties of Tb-Fe films, Thin Sol. Film 583 (2015) 1.

\bibitem{UCB_JMMM_2016}
 K. Umadevi, J. Arout Chelvane, H. Basumatary, M. Ramudu, S. V. Kamat, V. Jayalakshmi, Role of processing parameters on the morphology and magnetic properties of Tb-Fe-Co thin films, J. Magn. Magn. Mater. 418 (2016) 163.

\bibitem{TAM_AIP_2017}
 A. Talapatra, J. Arout Chelvane, J. Mohanty, Microscopic understanding of domain formation in Gd-Fe thin films, AIP Conf. Proc. 1832 (2017) 130044.

\bibitem{TAM_JMMM_2016}
 A. Talapatra, J. Mohanty, Laser induced local modification of magnetic domain in Co/Pt multilayer, J. Magn. Magn. Mater. 418 (2016) 224.

\bibitem{TAM_AIP_2016}
 A. Talapatra, J. Mohanty, Magnetic domain and domain wall in Co/Pt multilayer, AIP Conf. Proc. 1731 (2016) 130027.

\bibitem{LMV_SCI_14}
 C-H. Lambert, S. Mangin, B. S. D. Ch. S. Varaprasad, Y. K. Takahashi, M. Hehn, M. Cinchetti, G. Malinowski, K. Hono, Y. Fainman, M. Aeschlimann, E. E. Fullerton, All-optical control of ferromagnetic thin films and nanostructures, Science 345 (2014) 1337.

\bibitem{MMK_PRB_2012}
 A. Maziewski, P. Mazalski, Z. Kurant, M. O. Liedke, J. McCord, J. Fassbender, J. Ferr\'{e}, A. Mougin, A. Wawro, L. T. Baczewski, A. Rogalev, F. Wilhelm, T. Gemming, Tailoring of magnetism in Co/Pt/Co ultrathin films by ion irradiation, Phys. Rev. B 85 (2012) 054427.

\bibitem{SCL_PMC_2017}
 C. Song, B. Cui, F. Li, X. Zhou, F. Pan, Recent progress in voltage control of magnetism: materials, mechanisms, and performance, Prog. Mater. Sci. 87 (2017) 33.

\bibitem{CMF_JPD_2013}
 S. Chung, S. M. Mohseni, V. Fallahi, T. N. Anh Nguyen, N. Benatmane, R. K. Dumas, J. \AA{}kerman, Tunable spin configuration in [Co/Ni]-NiFe spring magnets, J. Phys. D: Appl. Phys. 46 (2013) 125004.

\bibitem{AMM_HI_2008}
 J. Arout Chelvane, G. Markandeyulu, M. Manivel Raja, Magnetic properties and M\"{o}ssbauer studies in $Y_{1-x}Gd_xFe_2$, Hyperfine Interact. 184 (2008) 27.

\bibitem{OOMMF}
 M. Donahue, D. G. Porter, OOMMF user's guide, version 1.0,
 Intergency Report NISTIR 6376, Nat. Inst. of Standard and Tech.
 Gaithersburg, MD, \url{http://math.nist.gov/oommf}.

 \bibitem{TAM_JMMM_2018}
 A. Talapatra, J. Arout Chelvane, J. Mohanty, Tuning magnetic microstructure in Gd-Fe thin films: experiment and simulation, J. Magn. Magn. Mater. 448 (2018) 360.

\end{thebibliography}

\end{document}